\documentclass[%
 reprint,
superscriptaddress,
 amsmath,amssymb,
 aps,
]{revtex4-1}

\usepackage{graphicx}
\usepackage{dcolumn}
\usepackage{bm}
\usepackage{hyperref}

\usepackage{color}
\usepackage{xcolor}

\begin{document}

\title{Probing bumblebee gravity with black hole X-ray data}

\author{Jiale~Gu}
\affiliation{Center for Field Theory and Particle Physics and Department of Physics, Fudan University, 2005 Songhu Road, 200438 Shanghai, China}

\author{Shafqat~Riaz}
\affiliation{Center for Field Theory and Particle Physics and Department of Physics, Fudan University, 2005 Songhu Road, 200438 Shanghai, China}

\author{Askar~B.~Abdikamalov}
\affiliation{Center for Field Theory and Particle Physics and Department of Physics, Fudan University, 2005 Songhu Road, 200438 Shanghai, China}
\affiliation{Ulugh Beg Astronomical Institute, Astronomy Str. 33, Tashkent 100052, Uzbekistan}
\affiliation{Institute of Fundamental and Applied Research, National Research University TIIAME, Kori Niyoziy 39, Tashkent 100000, Uzbekistan}

\author{Dimitry~Ayzenberg}
\affiliation{Theoretical Astrophysics, Eberhard-Karls Universität Tübingen, Auf der Morgenstelle 10, D-72076 Tübingen, Germany}

\author{Cosimo~Bambi}
\email[Corresponding author: ]{bambi@fudan.edu.cn}
\affiliation{Center for Field Theory and Particle Physics and Department of Physics, Fudan University, 2005 Songhu Road, 200438 Shanghai, China}

\date{\today}

\begin{abstract}
Bumblebee gravity is one of the simplest gravity theories with spontaneous Lorentz symmetry breaking. Since we know a rotating black hole solution in bumblebee gravity, we can potentially test this model with the available astrophysical observations of black holes. In this work, we construct a reflection model in bumblebee gravity and we use our model to analyze the reflection features of a \textsl{NuSTAR} spectrum of the Galactic black hole EXO~1846--031 in order to constrain the Lorentz-violating parameter $\ell$. We find that the analysis of the reflection features in the spectrum of EXO~1846--031 cannot constrain the parameter $\ell$ because of a very strong degeneracy between the estimates of $\ell$ and of the black hole spin parameter $a_*$. Such a degeneracy may be broken by combining other observations.
\end{abstract}

\maketitle


\section{Introduction\protect}\label{sec:intro}

General relativity was proposed by Einstein at the end of 1915~\cite{Einstein:1916vd} and has successfully survived until today without any modification. The predictions of general relativity have been extensively tested in the so-called weak field regime, mainly with Solar System experiments and radio observations of binary pulsars~\cite{Will:2014kxa}. While weak field tests are still an active line of research, today most of the efforts are focused to test the theory on large scales with cosmological observations~\cite{Koyama:2015vza,Ferreira:2019xrr} and in the strong field regime with data from compact astrophysical objects~\cite{Berti:2015itd,Bambi:2015kza,Yagi:2016jml}. From the theoretical point of view, there are certainly a few issues suggesting the existence of physics beyond general relativity: the presence of spacetime singularities in most physically relevant solutions of the Einstein Equations, the black hole information paradox in the black hole evaporation process, the search for a UV-complete quantum theory of gravity, etc. Today the search and the study of theories of gravity beyond general relativity is a very active research field.

Lorentz invariance is a fundamental ingredient both in general relativity and in the Standard Model of particle physics. In the case of general relativity, it is always possible to choose a locally inertial reference frame, where the non-gravitational laws of physics are those of special relativity and the Lorentz symmetry holds. However, in theories beyond general relativity the Lorentz symmetry may be violated at some level~\cite{Kostelecky:1988zi,Kostelecky:1989jw,Gambini:1998it,Ellis:1999uh,Carroll:2001ws}.

Bumblebee gravity is a simple model in which the Lorentz symmetry is spontaneously broken by the vacuum expectation value of a vector field~\cite{Kostelecky:1989jw}. In such a framework, an exact non-rotating and uncharged black hole solution was found in Ref.~\cite{Casana:2017jkc}. However, astrophysical objects have normally a non-vanishing spin angular momentum and observational tests with black holes usually require to know the rotating black hole solution. An exact rotating black hole solution was found in Ref.~\cite{Ding:2019mal}, where the authors explored the impact of the Lorentz-violating parameter $\ell$ on the shape of the black hole shadow. They did not report any analysis to constrain the parameter $\ell$ from the comparison of their theoretical predictions with the observations of the Event Horizon Telescope Collaboration of the supermassive black hole in M87$^*$~\cite{EventHorizonTelescope:2019dse}, but they argued that the value of the Lorentz-violating parameter $\ell$ may be measured by future observations of black hole shadows.

In the present manuscript, we consider the rotating black hole solution in bumblebee gravity found in Ref.~\cite{Ding:2019mal} and we study possible observational signatures in the reflection spectrum of accretion disks of black holes~\cite{Bambi:2020jpe}. X-ray reflection spectroscopy is potentially a powerful technique to probe the strong gravity region of accreting black holes, both stellar-mass and supermassive, and it has been already applied successfully to test conformal gravity~\cite{Zhou:2018bxk,Zhou:2019hqk}, Kaluza-Klein gravity~\cite{Zhu:2020cfn}, asymptotically safe quantum gravity~\cite{Zhou:2020eth}, and Einstein-Maxwell dilaton-axion gravity~\cite{Tripathi:2021rwb}. Depending on the specific black hole metric predicted by the theory, with the currently available data, X-ray reflection spectroscopy may provide comparable -- and sometimes even more stringent -- constraints than gravitational wave observations by \textsl{LIGO} and \textsl{Virgo}, and normally an order of magnitude stronger constraints than black hole imaging; see, e.g., Refs.~\cite{Tripathi:2020yts,Tripathi:2021rqs,Riaz:2022rlx}.

Our paper is organized as follows. In Section~\ref{BBG}, we briefly review bumblebee gravity and its rotating black hole solution found in Ref.~\cite{Ding:2019mal}. In Section~\ref{xray-ref-spec}, we construct our reflection model in bumblebee gravity. Section~\ref{s:sa} is devoted to the spectral analysis of a \textsl{NuSTAR} observation of the Galactic black hole EXO~1846--031. Summary and conclusions are reported in Section~\ref{s:c}. Throughout the manuscript, we employ natural units in which $c = G_{\rm N} = 1$ and a metric with signature $(-+++)$.


\section{Black holes in bumblebee gravity}\label{BBG}

Bumblebee gravity is a class of models in which a vector field acquires a non-vanishing vacuum expectation value, leading to a spontaneous breaking of the Lorentz symmetry~\cite{Kostelecky:1989jw,Bluhm:2004ep}. In this section, we briefly review bumblebee gravity and its rotating black hole solution found in Ref.~\cite{Ding:2019mal}.

\subsection{Bumblebee gravity}

We consider the bumblebee gravity model described by the following action~\cite{Casana:2017jkc,Ding:2019mal}:
\begin{equation}\label{eqn:100}
\begin{aligned}
    S=\int d^4 x \sqrt{-g} \Big[ & \frac{1}{16\pi} \left( R+\varrho B^\mu B^\nu R_{\mu\nu} \right) \\
    & -\frac{1}{4}B^{\mu\nu}B_{\mu\nu} - V(B^\mu) \Big] \, ,
\end{aligned}
\end{equation}
where $\varrho$ is a real coupling constant controlling the non-minimal gravity interaction to the bumblebee vector field $B^\mu$, $B_{\mu\nu}$ is the bumblebee field strength
\begin{align}
    B_{\mu\nu}=\partial_{\mu}B_{\nu}-\partial_{\nu}B_{\mu} \, ,
\end{align}
and $V(B^\mu)$ is a certain potential of the bumblebee vector field that is used to induce the violation of the Lorentz symmetry.

The potential $V(B^\mu)$ has the form
\begin{align}
    V = V (B^{\mu}B_{\mu}\pm b^2) \, ,
\end{align}
where $b^2$ is a real positive constant. The potential must have a minimum at $B^{\mu}B_{\mu}\pm b^2 = 0$. The bumblebee field gets a non-vanishing vacuum expectation value $\langle B^\mu \rangle = b^\mu$, where $b^\mu$ is a vector field of constant norm: $b^\mu b_\mu = \mp b^2$ ($b^\mu$ can be either timelike or spacelike).

From the action in Eq.~(\ref{eqn:100}), we get the following field equations for the gravity sector:
\begin{align}
    R_{\mu\nu}-\frac{1}{2}Rg_{\mu\nu}= 8\pi T_{\mu\nu}^B \, , 
\end{align}
where the energy-momentum tensor of bumblebee field, $T_{\mu\nu}^B$, is given by~\cite{Casana:2017jkc}
\begin{equation}
\begin{aligned}\label{eqn:4}
    T_{\mu\nu}^B=&B_{\mu\alpha}{B^{\alpha}}_{\nu}-\frac{1}{4} g_{\mu\nu} B^{\alpha\beta}B_{\alpha\beta}-g_{\mu\nu}V+2B_\mu B_\nu V' \\
    & +\frac{\varrho}{8\pi} \Big[ \frac{1}{2} g_{\mu\nu} B^{\alpha}B^{\beta}R_{\alpha\beta}-B_{\mu}B^{\alpha}R_{\alpha\nu}-B_{\nu}B^{\alpha}R_{\alpha\mu} \\
    & +\frac{1}{2}\nabla_{\alpha}\nabla_{\mu}(B^{\alpha}B_{\nu})+\frac{1}{2}\nabla_{\alpha}\nabla_{\nu}(B^{\alpha}B_{\mu}) \\
    & -\frac{1}{2}\nabla^2(B_{\mu}B_{\nu})-\frac{1}{2}g_{\mu\nu}\nabla_{\alpha}\nabla_{\beta}(B^{\alpha}B^{\beta})\Big] \, ,
\end{aligned}
\end{equation}
and $V'$ is 
\begin{align}
    V'=\frac{\partial V(x)}{\partial x}\vert_{x=B^{\mu}B_{\mu}\pm b^2} \, .
\end{align}

The field equations of the bumblebee field are
\begin{align}
    \nabla^\mu B_{\mu\nu} = 2 V' B_\nu - \frac{\varrho}{8\pi} B^\mu R_{\mu\nu} , 
\end{align}
but in what follows we will assume that the bumblebee field is frozen to its vacuum expectation value, namely $B^\mu = b^\mu$.

\subsection{Black hole solution}

Assuming that the bumblebee field gets a purely radial vacuum expectation value, namely $b_\mu = \left( 0 , b_r , 0 , 0 \right)$, in Ref.~\cite{Ding:2019mal} the authors obtained the following exact, rotating black hole solution in bumblebee gravity in Boyer-Lindquist coordinates:
\begin{equation}\label{eqn:6}
\begin{aligned}
 ds^2=&-\left( 1-\frac{2Mr}{\rho^2} \right) dt^2 - \frac{4Mra\sqrt{1+\ell} \sin^2\theta}{\rho^2} dtd\phi \\
 ~~& + \frac{\rho^2}{\Delta} dr^2 + \rho^2d\theta^2 + \frac{A\sin^2\theta}{\rho^2}d\phi^2 \, ,
\end{aligned}
\end{equation}
where
\begin{align}
\label{best-fit-model}
 &\rho^2 = r^2+ \left( 1+\ell \right) a^2 cos^2\theta \, ,~~~~\\ 
 &\Delta = \frac{r^2-2Mr}{1+\ell}+a^2 \, ,~~~~~\\
 &A = \left[r^2+ \left(1+\ell\right)a^2\right]^2-\Delta\left(1+\ell\right)^2 a^2 \sin^2\theta \, .
\end{align}
Here $M$ is the black hole mass, $a$ is the specific spin parameter (in what follows, we will often use the dimensionless spin parameter $a_* = a/M$), and $\ell$ is the Lorentz-violating parameter. For $\ell = 0$, we exactly recover the Kerr solution of general relativity and the Lorentz symmetry holds, while a non-vanishing value of $\ell$ leads to a deviation from the Kerr solution and indicates that the Lorentz symmetry is violated, since $\ell$ is proportional to the coupling constant $\varrho$ and the vacuum expectation value of the bumblebee field (for more details, see Refs.~\cite{Casana:2017jkc,Ding:2019mal}).

Like for the Kerr spacetime, the radial coordinate of the event horizon can be inferred from the equation $g^{rr}=0$ and turns out to be
\begin{align}
    r_{\rm H}=M + \sqrt{M^2-a^2(1+\ell)} \, ,
\end{align}
which requires
\begin{align}\label{eqn:11}
    |a| \leq \frac{M}{\sqrt{1+\ell}} \, .
\end{align}
If Eq.~(\ref{eqn:11}) is violated, there is no horizon and the spacetime has a naked singularity. In our analysis, we will not consider the possibility of a naked singularity and we will construct a reflection model only in the parameter space with black holes.

\begin{figure}[tbp]
\includegraphics[width=0.48\textwidth,trim=0.1cm 0.1cm 0.05cm 0.0cm,clip]{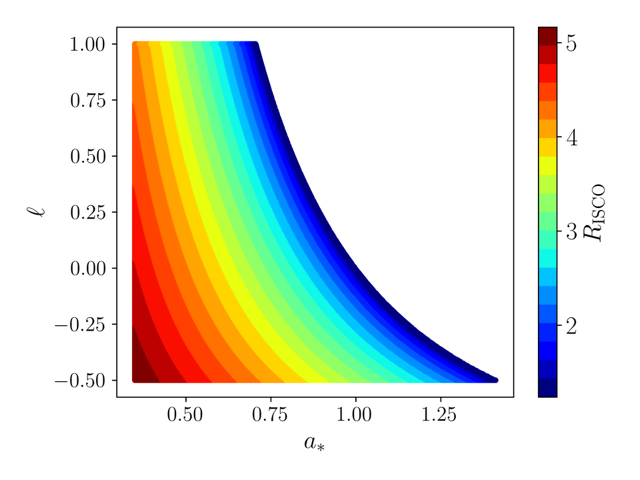}
\caption{\label{Risco PLOTS} 
Contour levels for the value of the ISCO radius $R_{\rm ISCO}$ in Boyer-Lindquist coordinates on the plane $a_*$ vs. $\ell$. The white region is ignored in our study because it is the region of the spacetimes with a naked singularity where Eq.~(\ref{eqn:11}) is violated.}
\end{figure}

Among the properties of the spacetime, the innermost stable circular orbit (ISCO) in the equatorial plane is particularly important for the structure of thin accretion disks and the corresponding X-ray spectrum of the black hole (for the calculation of the ISCO radius, see, e.g., Ref.~\cite{Bambi:2017khi}). Fig.~\ref{Risco PLOTS} shows the value of the ISCO radius around a black hole in bumblebee gravity as a function of the black hole spin parameter $a_*$ and of the Lorentz-violating parameter $\ell$. The white region is ignored in our study because it does not satisfy Eq.~(\ref{eqn:11}).

We note that in our analysis we consider rather a broader range of $\ell$. Solar System experiments provide quite a stringent constraint on Lorentz-violating parameter, $\ell < 10^{-13}$ (Cassini spacecraft)~\cite{Casana:2017jkc}. However, if we interpret the bumblebee model as an effective theory, weak-field and strong-field tests probe different regimes and weak-field constraints may not be valid in the strong field regime. In Ref.~\cite{Wang:2021gtd}, the authors constrain the Lorentz-violating parameter from the observed frequencies of the quasi-periodic oscillations (QPOs) of the Galactic black hole GRO~J1655--40 within the relativistic precession model. Their 1-$\sigma$ measurement is $\ell = -0.10_{-0.13}^{+0.17}$. In what follows, we will not impose any prior restriction on the parameter $\ell$.

\begin{figure}[tbp]
\includegraphics[width=0.48\textwidth,trim=0.0cm 3.0cm 0.0cm 0.0cm,clip]{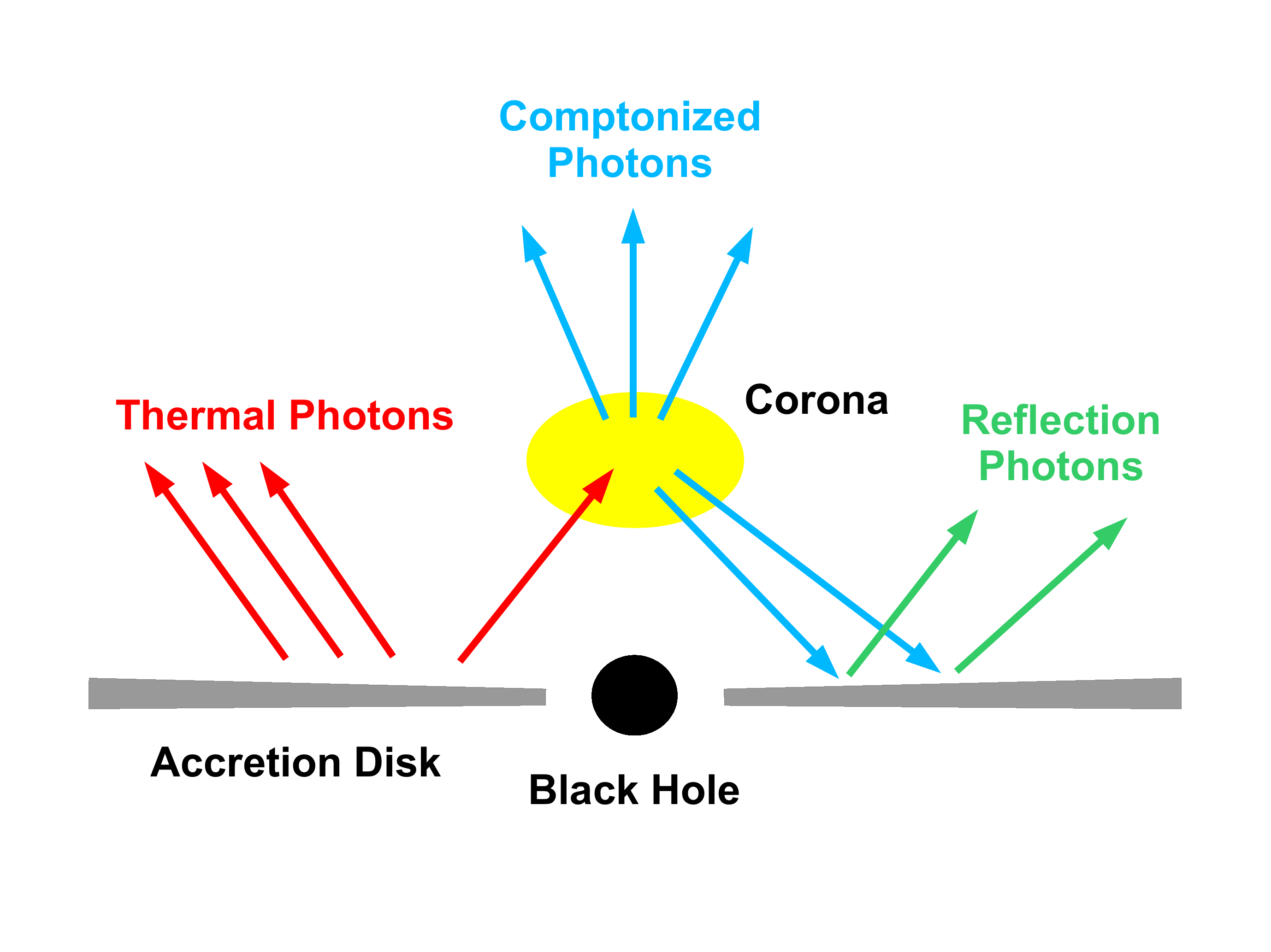}
\caption{\label{corona} Disk-corona model: the electromagnetic spectrum of the system is characterized by a multi-temperature blackbody component from the cold disk, a Comptonized spectrum from the hot corona, and a reflection spectrum from the disk. Figure from Ref.~\cite{Bambi:2021chr}. }
\end{figure}

\section{X-ray reflection spectroscopy}\label{xray-ref-spec}

Blurred reflection features are commonly observed in the X-ray spectra of accreting black holes and are thought to be generated by the illumination of a cold accretion disk by a hot corona (see, e.g., \cite{Bambi:2020jpe,Bambi:2017iyh} and references therein). The system with the main physical processes is illustrated in the cartoon in Fig.~\ref{corona}. The central black hole is accreting from a geometrically thin and optically thick accretion disk. Every point of the disk is in local thermal equilibrium and emits a blackbody spectrum, which becomes a multi-temperature blackbody spectrum when it is integrated radially. Such a thermal component is normally peaked in the soft X-ray band (0.1-1~keV) for stellar-mass black holes and in the UV band (1-100~eV) for supermassive black holes. The ``corona'' is some hotter plasma ($T_{\rm e} \sim 100$~keV) near the black hole and the inner part of the accretion disk, even if its exact morphology is not completely understood so far. The corona may be the atmosphere above the accretion disk, the accretion flow in the plunging region between the inner edge of the disk and the black hole, the base of the jet, etc. Thermal photons from the disk can inverse Compton scatter off free electrons in the corona. The spectrum of the Comptonized photons can be usually approximated by a power law component with a high-energy cutoff $E_{\rm cut} \approx 2$-3~$T_{\rm e}$. A fraction of the Comptonized photons can illuminate back the disk. Here we have Compton scattering and absorption followed by fluorescent emission, which produce the reflection spectrum.

In the rest-frame of the gas in the disk, the reflection spectrum is characterized by narrow fluorescent emission lines below 10~keV, the Fe K-edge at 7-10~keV, and the Compton hump peaked at 20-30~keV~\cite{Garcia:2013oma}. The iron K$\alpha$ line is often the most prominent line and it is at 6.4~keV in the case of neutral or weakly ionized iron atoms and can shift up to 6.97~keV in the case of H-like iron ions. In the reflection spectrum of the disk observed far from the source, these narrow fluorescent emission lines become broadened and skewed as a result of relativistic effects that affect the photons propagating from the emission point in the disk to the flat faraway region of the detector. In the presence of the correct astrophysical model, the analysis of these blurred reflection features can be a powerful tool to probe the strong gravitational field around black holes and test fundamental physics~\cite{Tripathi:2020yts,Tripathi:2021rqs,Bambi:2021chr,Tripathi:2018lhx}.

In order to test bumblebee gravity with black holes, we construct a reflection model employing the spacetime metric in Eq.~(\ref{eqn:6}). We implement such a black hole solution in {\tt relxill\_nk}~\cite{Bambi:2016sac,Abdikamalov:2019yrr,Abdikamalov:2020oci}, which is an extension of the {\tt relxill} package~\cite{Dauser:2013xv,Garcia:2013oma,Garcia:2013lxa} to non-Kerr spacetimes. The model employs the formalism of the Cunningham transfer function~\cite{Cunningham:1975zz}, where a transfer function is introduced to encode all the relativistic effects of the spacetime and all the assumptions about the structure of the accretion disk. In {\tt relxill\_nk}, the accretion disk is described by the Novikov-Thorne model~\cite{Novikov:1973kta,Page:1974he}. The calculations of the transfer function in a generic stationary, axisymmetric, and asymptotically flat spacetime with a Novikov-Thorne disk have been already presented many times in literature (see, e.g., Refs.~\cite{Bambi:2017khi,Bambi:2016sac}), so here we only recall the main passages.

The flux of the reflection spectrum as observed far from the source can be written as
\begin{align}\label{eq:fff}
    F_{\rm obs}(\nu_{\rm obs}) &= \int {I_{\rm obs}(\nu_{\rm obs},X,Y)d\Omega}\\
    &=\int g^3 {I_{\rm e}(\nu_{\rm e},r_{\rm e},\vartheta_{\rm e})d\Omega} \, ,
\end{align}
where $\nu_{\rm obs}$ and $\nu_{\rm e}$ are, respectively, the photon frequency measured at the detection point by the distant observer and that measured at the emission point in the rest-frame of the gas of the disk; $I_{\rm obs}$ and $I_{\rm e}$ are, respectively, the specific intensity of the radiation at the detection point and at the emission point; $X$ and $Y$ are the Cartesian coordinates in the observer's sky and $d\Omega = dX dY/D^2$ is the line element of the solid angle subtended by the image of the disk in the observer's sky, where $D$ is the distance of the observer from the source. $r_{\rm e}$ is the emission radius in the disk and $\vartheta_{\rm e}$ is the emission angle with respect to the normal to the disk measured in the rest-frame of the fluid. $g = \nu_{\rm obs}/\nu_{\rm e}$ is the redshift factor and $I_{\rm obs} = g^3 I_{\rm e}$ follows from Liouville's theorem~\cite{Lindquist:1966igj}.

Introducing the transfer function $f(g^*,r_{\rm e},i)$, Eq.~(\ref{eq:fff}) can be rewritten as
\begin{align}
    F_{\rm obs}(\nu_{\rm obs})=\int_{R_{\rm in}}^{R_{\rm out}} \int_{0}^{1} 
    \frac{\pi r_{\rm e} g^2 f(g^*,r_{\rm e},i)}{\sqrt{g^*(1-g^*)}} I_{\rm e}(r_{\rm e},\vartheta_{\rm e}) dg^* dr_{\rm e} \, ,
\end{align}
where $R_{\rm in}$ and $R_{\rm out}$ are, respectively, the inner and the outer edge of the disk, $i$ is the inclination angle of the disk with respect to the line of sight of the distant observer, $g^*$ is the relative redshift defined as
\begin{align}
    g^*=\frac{g-g_{\rm min}}{g_{\rm max}-g_{\rm min}} \, ,
\end{align}
and $g_{\rm max} = g_{\rm max} (r_{\rm e},i)$ and $g_{\rm min} = g_{\rm min} (r_{\rm e},i)$ are, respectively, the maximum and the minimum value of the redshift factor for the photons emitted at the radial coordinate $r_{\rm e}$ and detected by a distant observer with polar coordinate $i$. The expression of the transfer function is
\begin{align}
    f(g^*,r_{\rm e},i)=\frac{1}{\pi r_{\rm e}} g \sqrt{g^*(1-g^*)} \left|\frac{\partial(X,Y)}{\partial(g^*,r_{\rm e})}\right| \, ,
\end{align}
where $|\partial(X,Y)/\partial(g^*,r_{\rm e})|$ is the Jacobian between the variables in the plane of the distant observer and in the disk.

\begin{figure*}[tbp]
\includegraphics[width=0.90\textwidth,trim={0.0cm 0.0cm 0.0cm 0.0cm},clip]{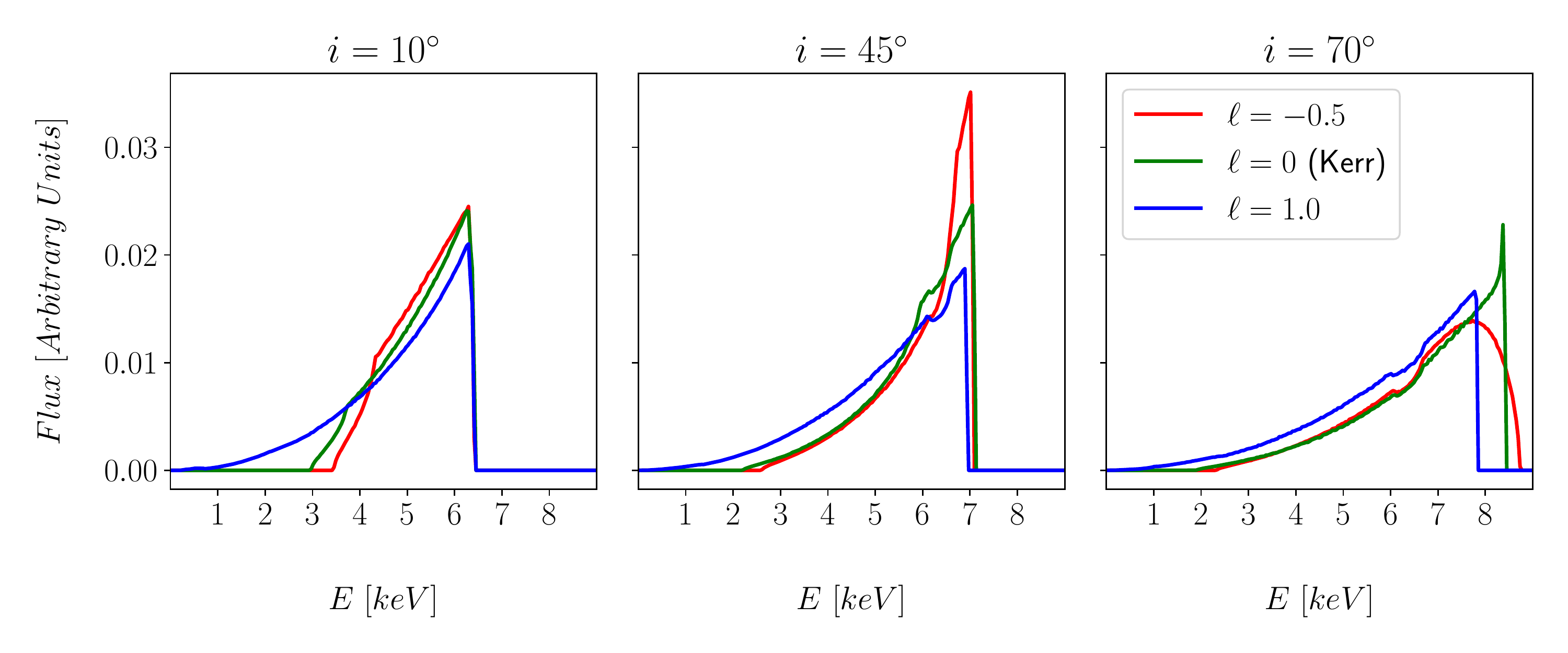}
\vspace{-0.5cm}
\caption{\label{ironlines}
Broadened iron lines from the accretion disk around a black hole in bumblebee gravity with spin parameter $a_{*}=0.7$, Lorentz-violating parameter $\ell = -0.5$ (red curves), 0 (green curves), and 1 (blue curves), and the inclination angle of the disk $i=10^{\circ}$ (left panel), $45^{\circ}$ (center panel), and $70^{\circ}$ (right panel). In these calculations, the inner edge of the disk is set at the ISCO radius, the outer edge at 400~$M$, and it is assumed an emissivity profile described by a power law with emissivity index 3; i.e., $I_{\rm e} \propto 1/r_{\rm e}^3$.}
\end{figure*}

The transfer function for a specific system characterized by the black hole spin parameter $a_*$, the Lorentz-violating parameter $\ell$, and the inclination angle of the disk $i$ is calculated by a ray-tracing code and tabulated in a FITS file in a grid of emission radii and relative redshifts $(r_{\rm e},g^*)$, as described in Refs.~\cite{Bambi:2016sac,Abdikamalov:2019yrr}. Assuming that the emission is a narrow iron line at 6.4~keV, our model predicts the broadened lines shown in Fig.~\ref{ironlines} for a black hole spin parameter $a_* = 0.7$ and different values of the Lorentz-violating parameter $\ell$ and of the inclination angle of the disk $i$. In the next section, we will assume that $I_{\rm e}$ is a reflection spectrum and we will use {\tt relxill\_nk} to analyze the reflection features of a specific observation of the Galactic black hole EXO~1846--031.


\section{Observational constraints}\label{s:sa}

In this section, we present the spectral analysis of a \textsl{NuSTAR} observation of the Galactic black hole EXO~1846--031 using the reflection model constructed in the previous section with the goal to constrain the value of the Lorentz-violating parameter $\ell$.

\subsection{Observation and data reduction}\label{obs-data-red}

EXO~1846--031 is a Galactic low-mass X-ray binary system~\cite{parmar1993discovery} discovered on April 3, 1985 by the \textsl{EXOSAT} mission~\cite{exosat}. After about 25~years in a quiescent state, on July 23, 2019, this source entered a new outburst, which was detected by the \textsl{MAXI} X-ray mission. On August 3, 2019, the X-ray mission \textsl{NuSTAR}~\cite{NuSTAR:2013yza} observed EXO~1846--031 under the observation ID 90501334002~\cite{Draghis:2020ukh}. The \textsl{NuSTAR} observation lasted 22.2~ks. In this observation, the source was very bright and its spectrum was simple and characterized by a very prominent and broadened iron line. Moreover, \textsl{NuSTAR} is the most suitable X-ray observatory available today for the analysis of reflection features of bright sources, thanks to its wide energy band capable of observing simultaneously both the fluorescent emission lines in the soft X-ray band and the Compton hump, as well as because the detectors are not affected by pile-up issues even in the case of bright sources. Previous analyses of this observation have also shown that the inner edge of the disk is very close to the black hole and illuminated well by the hot corona, which are two more factors that maximize the relativistic effects in the reflection features~\cite{Draghis:2020ukh,Tripathi:2020yts,Roy:2021pns}. All these ingredients make this specific \textsl{NuSTAR} spectrum particularly suitable for testing the Kerr hypothesis with {\tt relxill\_nk}~\cite{Tripathi:2020yts}.

We follow the data reduction method described in Ref.~\cite{Draghis:2020ukh}. Using the HEASOFT module {\tt nupipeline} and the calibration database CALDB 20220301, we cleaned the raw data and generated an event file. In both the Focal Plane Module (FPM) A and B sensors, the source spectra were collected from a 180-arcsecond-radius region. To generate the background spectra, we chose the same region as the source's size but far away to eliminate the influence of the source's photons. The response matrix file, the auxiliary file, and the source and background spectra were all made with the help of the HEASOFT module {\tt nuproducts}. The spectra were then grouped using {\tt grppha} to ensure that there were at least 30 counts per energy bin. We utilized the updated calibration file, therefore we did not need to include the {\tt nuMLIv1.mod} table in the fitting model to tackle the issue in the 3-7~keV flux region caused by a tear in the thermal blanket of FPMA~\cite{Draghis:2020ukh}.

\begin{table}[tbh]
\caption{The best-fit values from the analysis of the 2019 \textsl{NuSTAR} data of EXO~1846--031. The reported uncertainties correspond to the 90\% confidence level for one relevant parameter ($\Delta \chi^2 = 2.71$). 
B means that the 90\% confidence level limit is not in the parameter range allowed by the model or the best-fit value is stuck at a boundary of the range allowed by the model: $q_{\rm in}$ and $q_{\rm mid}$ are allowed to vary in the range $(0,10)$, $r_{\rm br2}$ is allowed to vary in the range $(R_{\rm ISCO}, 900~M)$, and $a_*$ and $\ell$ are allowed to vary in the parameter space shown in Fig.~\ref{Risco PLOTS} and Fig.~\ref{constraint}.}
\vspace{0.4cm}
\label{t-fit}
{\renewcommand{\arraystretch}{1.3}
\begin{tabular}{lccccc}
\hline\hline
{\tt tbabs} &&&&& \\
$N_{\rm H}$ [$10^{22}$ cm$^{-2}$] & $$ &&& $$ & $6.5^{+0.7}_{-0.3}$ \\
\hline
{\tt diskbb}&&&&& \\
$kT_{\rm in}$ [keV] & $$ &&& $$ & $0.317^{+0.067}_{-0.008}$ \\
norm [$10^{3}$] & $$ & && $$ & $100^{+63}_{-68}$ \\
\hline
{\tt cutoffpl} \\
$\Gamma$ & $$ &&& $$ & $2.04^{+0.06}_{-0.06}$ \\
$E_{\rm cut}$ [keV] & $$ &&& $$ & $112^{+13}_{-6}$ \\
norm & $$ &&& $$ & $2.04^{+0.08}_{-0.08}$ \\
\hline
{\tt relxillion\_nk} &&&&& \\
$q_{\rm in}$ & $$ &&& $$ & $100_{-0.7}^{\rm +B}$ \\
$r_{\rm br1}$ [$r_{\rm g}$] & $$ &&& $$ & $5.7^{+2.3}_{-1.9}$ \\
$q_{\rm mid}$ & $$ &&& $$ & $0.2^{+0.8}_{\rm +B}$ \\
$r_{\rm br2}$ [$r_{\rm g}$]  & $$ &&& $$ & $352^{\rm +B}_{-133}$ \\
$q_{\rm out}$ & $$ &&& $$ & $3^*$ \\
$i$ [deg] & $$ &&& $$ & $83.2^{+0.8}_{-1.0}$ \\
$a_*$ & $$ &&& $$ & $0.82^{\rm +B}_{\rm -B}$ \\
$l$   & $$ &&& $$ & $0.46^{\rm +B}_{\rm -B}$ \\
$\log\xi_{\rm in}$ [erg~cm~s$^{-1}$] & $$ &&& $$ & $3.00^{+0.04}_{-0.12}$ \\
$\alpha_{\xi}$ & $$ &&& $$ & $0.19^{+0.05}_{-0.04}$\\
$A_{\rm Fe}$ & $$ &&& $$ & $1.5^{+0.4}_{-0.5}$ \\
norm [$10^{-3}$] & $$ & && $$ & $6.2^{+0.5}_{-0.3}$ \\
\hline
$\chi^2/\nu$ &&&&& $\quad 2658.33/2598 \quad$ \\
&&&&& =1.02322 \\
\hline\hline
\end{tabular}}
\end{table}

\subsection{Spectral analysis}\label{spec-analysis}

We used XSPEC 12.10.1s~\cite{xspec} with WILMS~\cite{Wilms:2000ez} abundances and VERNER cross-section to analyze the data. Initially, we fit the data with an absorbed power-law model ({\tt tbabs*cutoffpl} in XSPEC language). {\tt tbabs} is responsible for modeling the Galactic absorption and has a free parameter, the hydrogen column density ($N_{\rm H}$)~\cite{Wilms:2000ez}. {\tt cutoffpl} models the power-law component of the corona using three free parameters: the photon index ($\Gamma$), the high-energy cutoff ($E_{\rm cut}$), and a normalization. We employed a floating constant between the FPMA and FPMB detectors to adjust for their slight difference. The initial fit residuals reveal a disk thermal component below 4~keV, a very broadened iron line at 5-7~keV, and a Compton hump at 10-30~keV. The latter two are the well-known relativistic reflection features already discussed in Section~\ref{xray-ref-spec}. Therefore, we added {\tt diskbb}~\cite{Makishima:1986ap} and {\tt relxillion\_nk}~\cite{Abdikamalov:2021rty}, respectively, to model the disk's thermal component and the relativistic reflection features. We can write the full model as
\begin{equation}
    \begin{aligned}
\label{best-fit-model}
   {\tt const*tbabs*(cutoffpl + diskbb }\\+ {\tt relxillion\_nk).}
\end{aligned}
\end{equation}

{\tt diskbb} is a relatively simple model, employing a Newtonian accretion disk, and is fully described by two parameters: the temperature at the inner edge of the disk $(T_{\rm in})$ and the normalization of the component. We chose {\tt diskbb} instead of a more sophisticated relativistic model like {\tt nkbb}~\cite{Zhou:2019fcg} because the thermal component in this observation is very weak and we do not have reliable measurements of the black hole mass and distance, which are instead required in any relativistic model for the thermal spectrum of the disk.

In our analysis, we used {\tt relxillion\_nk}, which is the version of {\tt relxill\_nk} with a non-vanishing disk's ionization gradient~\cite{Abdikamalov:2019yrr}. Here the ionization parameter $\xi$ is not constant over the radial coordinate but it is given by
\begin{align}
    \xi(r) = \xi_{\rm in} \bigg ( \frac{R_{\rm in}}{r} \bigg)^{\alpha_{\xi}} \, , 
\end{align}
where $\xi_{\rm in}$ is the value of the ionization parameter at the inner edge of the disk and $\alpha_{\xi}$ is the ionization index, which we kept free in the fitting process. The data clearly prefer a non-trivial ionization profile. {\tt relxillion\_nk} can fit well an absorption feature around 7~keV, which would otherwise require a Gaussian component in the model~\cite{Draghis:2020ukh}. Without a non-vanishing ionization gradient, we cannot constrain well the normalization of {\tt cutoffpl}, while we can do it with a free ionization index. The difference of the minimum of $\chi^2$ between without and with ionization gradient is around 40.

Concerning the emissivity profile, we initially fit the data assuming a broken power-law, and we found a very steep inner emissivity profile ($q_{\rm in} = 10$) and a nearly flat outer emissivity profile ($q_{\rm out} \approx 0$). Such an emissivity profile may resemble an extended and uniform corona, which is not a self-consistent model because the luminosity of the disk's outer part should scale with $r^2/r^{q_{\rm out}}$, hence $q_{\rm out}$ must be greater than 2 for the disk to have a finite luminosity. So, we re-fit the data with a twice broken power-law with four free parameters: the inner emissivity index $q_{\rm in}$, the breaking radius between the disk's inner and central parts $r_{\rm br1}$, the central emissivity index $q_{\rm mid}$, and the breaking radius between the disk's central and outer parts $r_{\rm br2}$. The outer emissivity index, $q_{\rm out}$, is fixed at 3, the typical value at larger radii for most coronal geometries. We fix the disk's inner and outer edge parameters to the ISCO and 900~$M$, respectively.

\begin{figure}[tbp]
\includegraphics[width=0.48\textwidth,trim={1.0cm 1.2cm 2.0cm 2.0cm},clip]{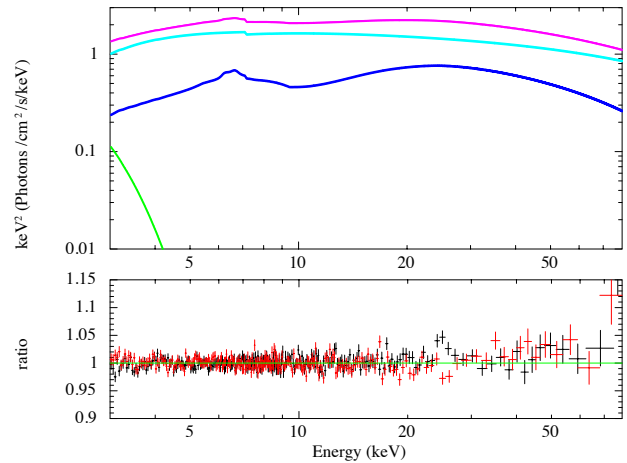}
\caption{\label{ratio} The upper quadrant shows the total best-fit model in magenta with the individual components: the thermal spectrum ({\tt diskbb}) is in green, the relativistic reflection spectrum ({\tt relxillion\_nk}) is in blue, and the Comptonized spectrum from the corona ({\tt cutoffpl}) is in cyan. The lower quadrant shows the data-to-best-fit model ratio, where the red and black colors represent the data from the FPMA and FPMB sensors, respectively.}
\end{figure}

\begin{figure}[tbp]
\includegraphics[width=0.48\textwidth,trim={0.4cm 0.8cm 0cm 0cm},clip]{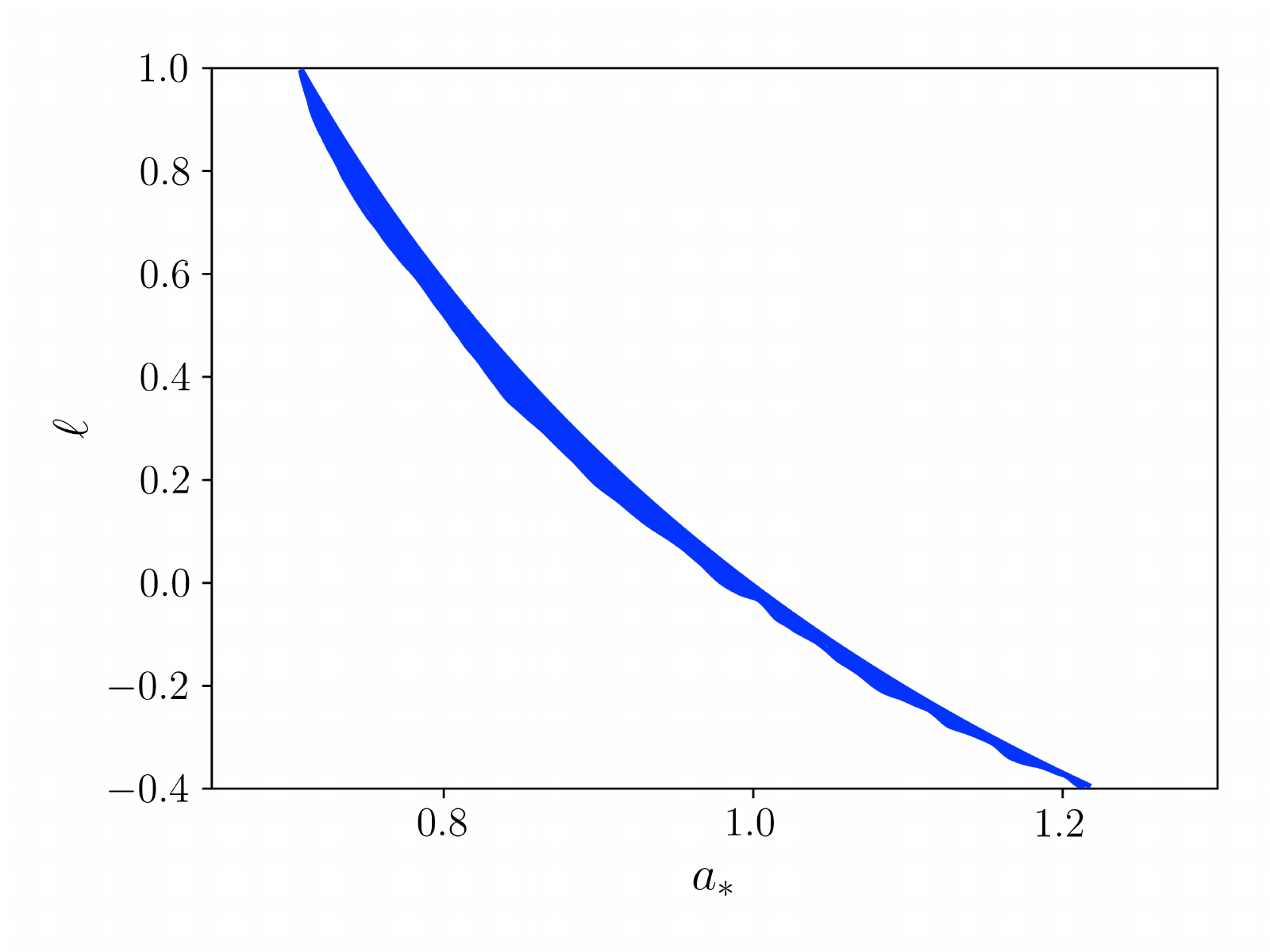}
\caption{\label{constraint} Constraints on the spacetime parameters $a_*$ and $\ell$ from our analysis of the 2019 \textsl{NuSTAR} data of EXO~1846--031. The blue-filled region represents the 99\% confidence level area for the two relevant parameters and the white region shows the least probable region. We clearly see a very strong degeneracy between $a_*$ and $\ell$.}
\end{figure}

We report the parameter estimate of the best-fit model in Table~\ref{t-fit}. We show the best-fit model with individual components and the data to best-fit model ratio in Fig.~\ref{ratio}. The constraints on the spin parameter $a_*$ and the Lorentz-violating parameter $\ell$ are shown in Fig.~\ref{constraint}. The blue-filled region in Fig.~\ref{constraint} corresponds to the 99\% confidence level area for the two relevant parameters. As we can see, there is a strong degeneracy between these two parameters and we cannot constrain $\ell$. We postpone the discussion of our result to the next section.

\section{Concluding remarks}\label{s:c}

In this work, we have presented the construction of a reflection model in bumblebee gravity and the spectral analysis of the 2019 \textsl{NuSTAR} observation of the Galactic black hole EXO~1846--031. The bumblebee gravity model has an extra parameter with respect to the general relativity one, the Lorentz-violating parameter $\ell$. For $\ell = 0$, the vacuum expectation value of the bumblebee vector field vanishes and we recover general relativity, where uncharged black holes are described by the Kerr solution. The possibility of a non-vanishing value of $\ell$ would instead indicate a deviation from the Kerr solution and a violation of Lorentz invariance.

The analysis of the 2019 \textsl{NuSTAR} observation of EXO~1846--031 is not new, and the discussion about the astrophysical properties of the system can be found in previous papers in the literature~\cite{Draghis:2020ukh,Tripathi:2020yts,Abdikamalov:2019yrr}. Here we focus the discussion only on the estimate of the Lorentz-violating parameter $\ell$.

Our result is summarized in Fig.~\ref{constraint}, where we see a very strong degeneracy between the black hole spin parameter $a_*$ and the Lorentz-violating parameter $\ell$. As a result of this degeneracy, we cannot constrain the Lorentz-violating parameter $\ell$ without an independent estimate of the black hole spin. Fig.~\ref{constraint} should also be compared with Fig.~\ref{Risco PLOTS}, and we can immediately conclude that the constraint on the plane $a_*$ vs $\ell$ follows the contour levels of the ISCO radius in Fig.~\ref{Risco PLOTS}. While the degeneracy of the ISCO radius between the black hole spin and the parameter quantifying the deviations from general relativity is the typical situation in models beyond general relativity, so Fig.~\ref{Risco PLOTS} is not a surprise, the rich structure of the reflection spectrum can normally break such a degeneracy and the analysis of the reflection features of fast-rotating black holes with the inner edge of their accretion disk very close to the black hole horizon can normally constrain the deformation parameter of the model; see, e.g., the results for specific theories of gravity beyond general relativity in Refs.~\cite{Zhou:2018bxk,Zhou:2019hqk,Zhu:2020cfn,Zhou:2020eth,Tripathi:2021rwb} and for phenomenological deformation parameters in Refs.~\cite{Tripathi:2020yts,Tripathi:2021rqs}.

In order to break the parameter degeneracy shown in Fig.~\ref{constraint}, it would be necessary to add constraints on $a_*$ and/or $\ell$ from other observations. Since the thermal spectrum of the disk is also determined by the background metric, we may think that the analysis of a spectrum showing simultaneously a strong thermal component and prominent blurred reflection features can do the job and break the parameter degeneracy. Similar spectra have been analyzed and are those that provide the strongest constraints on the parameter quantifying the deviations from general relativity~\cite{Tripathi:2021rqs,Tripathi:2020dni,Zhang:2021ymo}. However, such a solution can unlikely work in this case of black holes in bumblebee gravity because the thermal component is normally degenerate with respect to the inner edge of the disk, so the same degeneracy found in Fig.~\ref{constraint} between $a_*$ and $\ell$. It is possible that other methods, like the measurement of the frequency of quasi-periodic oscillations~\cite{Bambi:2013fea} or the estimate of the jet power~\cite{Bambi:2012ku,Bambi:2012zg}, can help to break the parameter degeneracy in the constraints on $a_*$ and $\ell$ found by using X-ray reflection spectroscopy, but the mechanisms behind those phenomena are not yet well understood, and current constraints would be highly model-dependent just because there are several models proposed in the literature and we do not know which of them, if any, is the correct one. The constraint on $\ell$ shown in Fig.~\ref{constraint} is probably close to the best we can do now with the available electromagnetic data of black holes.


\vspace{0.5cm}

{\bf Data Availability Statement.} We analyzed publicly available data that can be downloaded for free, for example,
from the HEASARC website \url{https://heasarc.gsfc.nasa.gov/}.

\begin{acknowledgments}
This work was supported by the Natural Science Foundation of Shanghai, Grant No. 22ZR1403400, the National Natural Science Foundation of China (NSFC), Grant No. 11973019, the Shanghai Municipal Education Commission, Grant No. 2019-01-07-00-07-E00035, and Fudan University, Grant No. JIH1512604.
\end{acknowledgments}

\end{document}